\documentclass[a4paper]{jpconf}

\usepackage{setstack}
\usepackage{graphicx}
\newcommand{\eg}{{\it e.g.}}

\newcommand{\pri}{^{\prime}}

\newcommand{\id}{\hat{1}}
\renewcommand{\Re}{\mathop{\mathrm{Re}}}
\renewcommand{\Im}{\mathop{\mathrm{Im}}}
\begin{document}

\title{Spin-transfer torque in magnetic multilayer nanopillars}

\author{J Peguiron$^1$, M S Choi$^2$ and C Bruder$^1$}

\address{$^1$ Department of Physics and Astronomy, University of Basel, Switzerland}
\address{$^2$ Department of Physics, Korea University, Seoul, South Korea}

\ead{joel.peguiron@unibas.ch}

\begin{abstract}
We consider a quasi one-dimensional configuration consisting of two small pieces of
ferromagnetic material separated by a metallic one and contacted by two metallic
leads. A spin-polarized current is injected from one lead. Our goal is to investigate the
correlation induced between the magnetizations of the two ferromagnets by spin-transfer
torque. This torque results from the interaction between the magnetizations
and the spin polarization of the current. We discuss the dynamics of a single ferromagnet,
the extension to the case of two ferromagnets, and give some estimates for
the parameters based on experiments.
\end{abstract}

\section{Introduction}

As proposed by Slonczewski~\cite{SloJMMM96} and Berger~\cite{BerPRB96}, a spin-polarized current flowing through a ferromagnetic layer exerts a torque on its magnetization. This torque, know as spin-transfer torque, can move the magnetization if the ferromagnet is small enough, as demonstrated in experiments~\cite{MyeSci99,KriSci05}. This mechanism allows for current-driven manipulation of the magnetization as an alternative to the manipulation with a magnetic field. Triggered by the quasi-classical model with spin-dependent potentials first presented in~\cite{SloJMMM96}, efforts have be made to refine the theoretical description of this mechanism~\cite{WaiPRB00,TseRMP05}. The resulting magnetization dynamics has been investigated \eg\ in~\cite{SloJMMM96,TseRMP05,SunPRB00,GmiPRL06}. In the present work, we evaluate the spin-transfer torque within a scattering approach, considering a single conducting channel in the ballistic regime. Interestingly, this approach yields an expression more general than the quasi-classical model used in~\cite{SloJMMM96} or the Landauer-B\"uttiker-like formalism developed in~\cite{WaiPRB00}, but similar to the results obtained within magnetoelectronic circuit theory~\cite{KovPRB06} or a diffusive transport analysis~\cite{GmiPRL06}. We also discuss the application of this approach to investigate the correlated dynamics of the magnetizations in a multilayer structure with two ferromagnetic layers. 

\section{Spin-transfer torque in a single ferromagnet}\label{sec-STT1FM}

\begin{figure}[b]
\begin{center}
\begin{minipage}[b]{2in}
\includegraphics[width=2in]{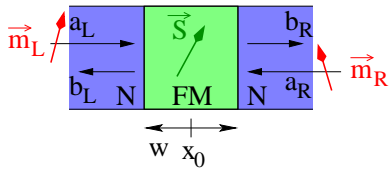}
\end{minipage}
\hspace{.2in}
\begin{minipage}[b]{4in}
\caption{\label{fig-1FMgen}Building block for a magnetic multilayer nanopillar: a small piece of ferromagnet between two normal metal areas. Currents with spin polarizations along~$\vec{m}_\mathrm{L,R}$ are injected from both sides (amplitudes~$a_\mathrm{L,R}$) and are partly transmitted to the other side, partly reflected to the same side (amplitudes~$b_\mathrm{L,R}$). The ferromagnetic layer of width~$w$ centered at position~$x_0$ carries a magnetization~$\vec{S}$.}
\end{minipage}
\end{center}
\end{figure}

We consider a small piece of ferromagnet contacted by two metallic leads such as in \fref{fig-1FMgen}.
%
%
We are interested in the dynamics of the magnetization~$\vec{S}$ of the ferromagnet caused by the flow of spin-polarized currents through the structure. By conservation of angular momentum, the change rate of the magnetization is given by the net spin flux transfered to the ferromagnet,~$\rmd\vec{S}/\rmd t=\vec{F}_\mathrm{net}$. The net spin flux is given by the difference of the rightward spin current densities at both sides of the ferromagnet,~$\vec{F}_\mathrm{net}=\vec{Q}_\mathrm{L}-\vec{Q}_\mathrm{R}$. This flux results from the change of spin polarization of the transmitted, reflected, and incoming currents caused by spin filtering at the metal-ferromagnet interfaces. This quantity is also called spin-transfer torque and reflects the reaction of the spin-polarized currents on the magnetization.

\begin{figure}
\begin{center}
\includegraphics[width=3in]{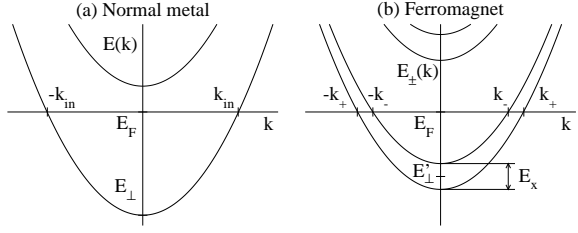}
\end{center}
\caption{\label{fig-BS}Band structure in (a) the normal metal areas and (b) the ferromagnet, and corresponding wave vectors at the Fermi level. We assume that a single channel with parabolic dispersion crosses the Fermi level. In the ferromagnet, the dispersions for the two spin polarizations with respect to the magnetization are shifted by the exchange energy~$E_\mathrm{X}$. We also allow for an overall shift~$E_\perp\pri-E_\perp$ of the band bottom with respect to the normal metal area.}
\end{figure}

For simplicity, we consider a single conducting channel below the Fermi level~[see \fref{fig-BS}(a)]. We model the propagation of the electrons along the structure by plane waves. The wave functions in the left and right metallic areas are thus given by
\begin{equation}
\eqalign{\Psi_\mathrm{L}(x)&
=\chi_\mathrm{aL}\rme^{\rmi k_\mathrm{in} x}+\chi_\mathrm{bL}\rme^{-\rmi k_\mathrm{in} x},\cr
\Psi_\mathrm{R}(x)&
=\chi_\mathrm{aR}\rme^{-\rmi k_\mathrm{in} x}+\chi_\mathrm{bR}\rme^{\rmi k_\mathrm{in} x},}
\end{equation}
where~$x$ is the coordinate along the structure and~$k_\mathrm{in}$ the Fermi wave vector. The spinor~$\chi$ of each wave gives the amplitudes of its two spin components. The corresponding spin current densities read
\begin{equation}\label{Eq-DefQ}
\eqalign{\vec{Q}_\mathrm{L}&=\frac{\hbar}{2}v_\mathrm{in}\left[\chi_\mathrm{aL}^\dagger\hat{\vec{\sigma}}\chi_\mathrm{aL}-\chi_\mathrm{bL}^\dagger\hat{\vec{\sigma}}\chi_\mathrm{bL}\right],\cr
\vec{Q}_\mathrm{R}&=\frac{\hbar}{2}v_\mathrm{in}\left[\chi_\mathrm{bR}^\dagger\hat{\vec{\sigma}}\chi_\mathrm{bR}-\chi_\mathrm{aR}^\dagger\hat{\vec{\sigma}}\chi_\mathrm{aR}\right],}
\end{equation}
where~$v_\mathrm{in}$ denotes the Fermi velocity and~$\hat{\vec{\sigma}}$ the vector of Pauli matrices. The spinors of the outgoing waves (denoted by an index~$a$) are related to the spinors of the ingoing ones (index~$b$) through the $4\times4$~scattering matrix~$\mathcal{S}$,
\begin{equation}\label{Eq-RelIO}
\left(\begin{array}{c}
\chi_\mathrm{bL}\\
\chi_\mathrm{bR}
\end{array}\right)
=\mathcal{S}\left(\begin{array}{c}
\chi_\mathrm{aL}\\
\chi_\mathrm{aR}
\end{array}\right). 
\end{equation}
Thus, it suffices to solve the scattering problem in order to obtain the net spin flux in terms of the spin polarization of the incoming currents.

In the ferromagnet, the electrons with spin parallel and anti-parallel to the magnetization have energies split by the exchange energy~$E_\mathrm{X}$~[see \fref{fig-BS}(b)].
%
%
As a consequence, there are two different wave vectors~$k_\pm$ at the Fermi energy~$E_\mathrm{F}$. Furthermore, the lateral confinement energy defining the band bottom may be different in the ferromagnet~($E_\perp\pri$) and in the normal metal~($E_\perp$). These two effects may be captured in a dimensionless exchange splitting~$\varepsilon=E_\mathrm{X}/(E_\mathrm{F}-E_\perp)$ and a dimensionless band shift $\Delta=(E^\prime_\perp-E_\perp)/(E_\mathrm{F}-E_\perp)$, yielding the wave vectors~$\kappa_\pm=k_\pm/k_\mathrm{in}=\sqrt{1-\Delta\pm\varepsilon/2}$ with respect to the Fermi wave vector~$k_\mathrm{in}$ in the normal metal.

Solving the matching equations for the wave function and its derivative at both interfaces, one obtains the scattering matrix
\begin{equation}
\mathcal{S}=\left(\begin{array}{cc}
\hat{r}\rme^{-\rmi k_\mathrm{in} (w-2x_0)} & \hat{t}\rme^{-\rmi k_\mathrm{in} w}\\
\hat{t}\rme^{-\rmi k_\mathrm{in} w} & \hat{r}\rme^{-\rmi k_\mathrm{in} (w+2x_0)}
\end{array}\right)
\end{equation}
for a ferromagnet of width~$w$ centered at position~$x_0$. It involves the transmission and reflection matrices~$\hat{t}=(\id+\vec{n}\cdot\hat{\vec{\sigma}})t_+/2+(\id-\vec{n}\cdot\hat{\vec{\sigma}})t_-/2$ and~$\hat{r}=(\id+\vec{n}\cdot\hat{\vec{\sigma}})r_+/2+(\id-\vec{n}\cdot\hat{\vec{\sigma}})r_-/2$, defined in terms of the unit vector~$\vec{n}=\vec{S}/S$ pointing in the direction of the magnetization, and of the transmission and reflection amplitudes (with the index~$\sigma=\pm$)
\begin{equation}
\eqalign{t_\sigma&=\left[\cos{(k_\sigma w)}-\frac{\rmi}{2}\left(\kappa_\sigma+\frac{1}{\kappa_\sigma}\right)\sin{(k_\sigma w)}\right]^{-1},\cr
r_\sigma&=\frac{\rmi}{2}\left(\kappa_\sigma-\frac{1}{\kappa_\sigma}\right)\sin{(k_\sigma w)}\left[\cos{(k_\sigma w)}-\frac{\rmi}{2}\left(\kappa_\sigma+\frac{1}{\kappa_\sigma}\right)\sin{(k_\sigma w)}\right]^{-1}.}
\end{equation}

Combining these results with~\eref{Eq-DefQ} and~\eref{Eq-RelIO}, one obtains an expression for the net spin flux transfered to the ferromagnet

\begin{equation}\label{Eq-Fnet}
\eqalign{
\vec{F}_\mathrm{net}=\frac{\hbar}{2}v_\mathrm{in}\ \vec{n}\times\Bigl\lbrack
&-\left(1-\Re\left\lbrace t_+^*t_-+r_+^*r_-\right\rbrace\right)\vec{n}\times\left(\vec{m}_\mathrm{L}+\vec{m}_\mathrm{R}\right)
-\Im\left\lbrace t_+^*t_-+r_+^*r_-\right\rbrace\left(\vec{m}_\mathrm{L}+\vec{m}_\mathrm{R}\right)\cr
&+\Re\left\lbrace t_+^*r_-+r_+^*t_-\right\rbrace\vec{n}\times\vec{m}_\mathrm{LR}-\Im\left\lbrace t_+^*r_-+r_+^*t_-\right\rbrace\vec{m}_\mathrm{LR}\Bigr\rbrack.}
\end{equation}
It involves the vectors~$\vec{m}_\mathrm{L}=\chi_\mathrm{aL}^\dagger\hat{\vec{\sigma}}\chi_\mathrm{aL}$ and~$\vec{m}_\mathrm{R}=\chi_\mathrm{aR}^\dagger\hat{\vec{\sigma}}\chi_\mathrm{aR}$, which point in the direction of the spin polarization of the currents incoming from the left and right metallic areas~[see~\eref{Eq-DefQ}], as well as the vector~$\vec{m}_\mathrm{LR}=2\Re\left\lbrace\rme^{-2\rmi k_\mathrm{in}x_0}\chi_\mathrm{aL}^\dagger\hat{\vec{\sigma}}\chi_\mathrm{aR}\right\rbrace$.

%
%
\begin{figure}[b]
\begin{center}
\begin{minipage}[b]{2in}
(a)\\
\includegraphics[width=2in]{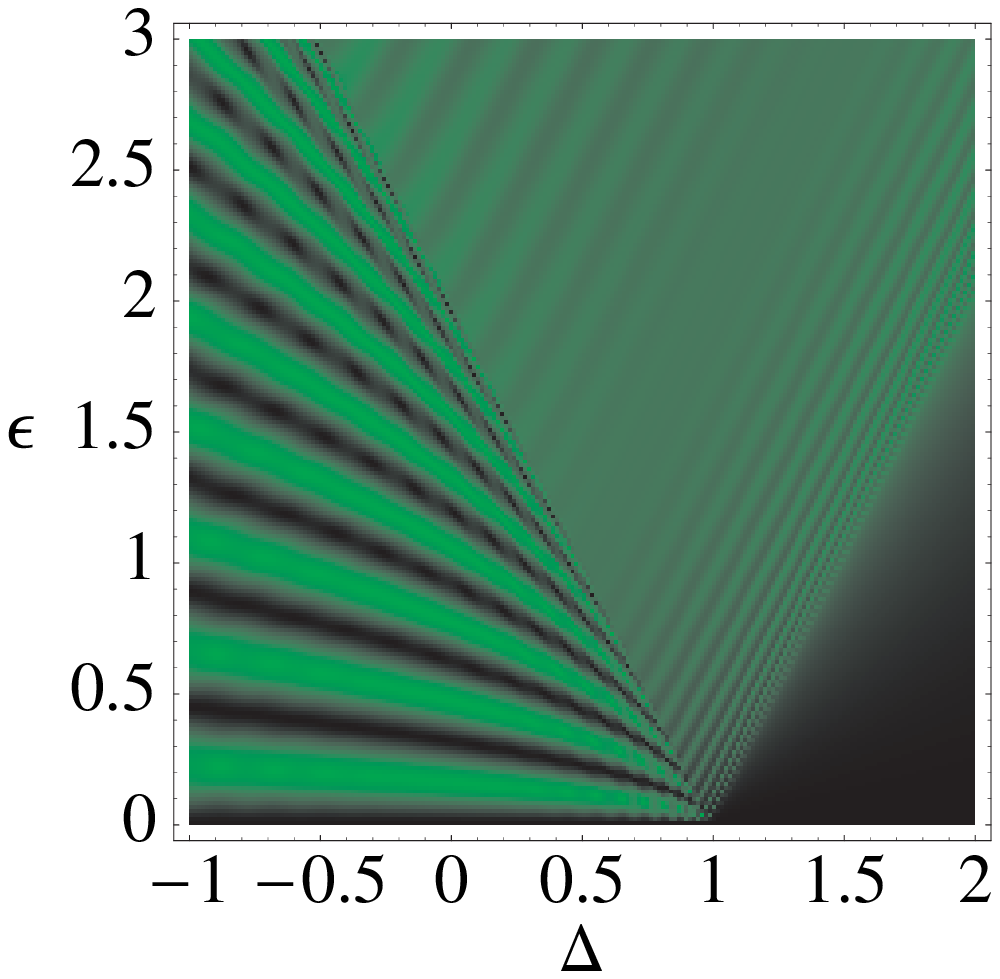}
(b)\\
\includegraphics[width=2in]{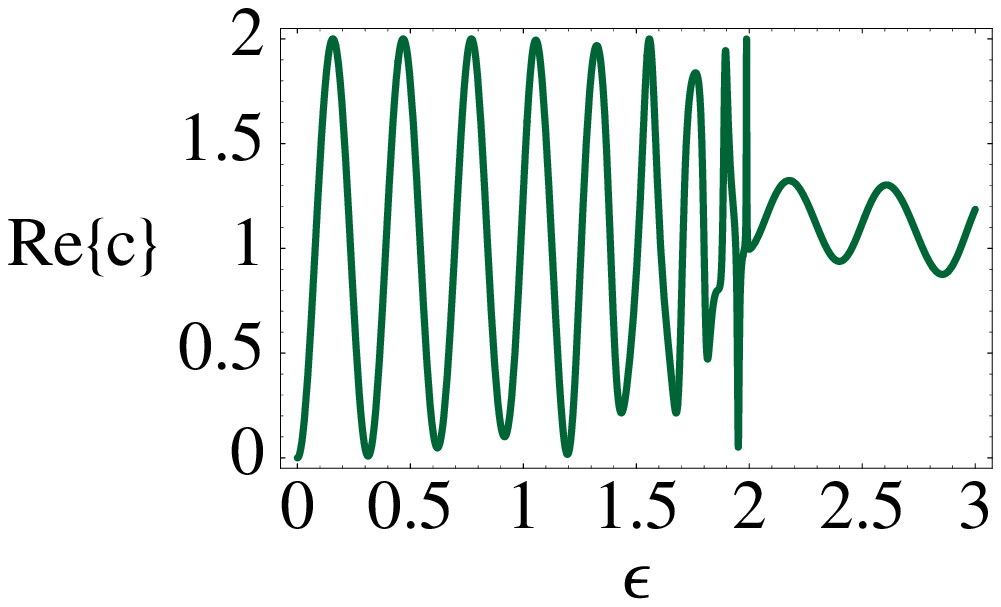}
\end{minipage}\hspace{.2in}
\begin{minipage}[b]{2in}
(c)\\
\includegraphics[width=2in]{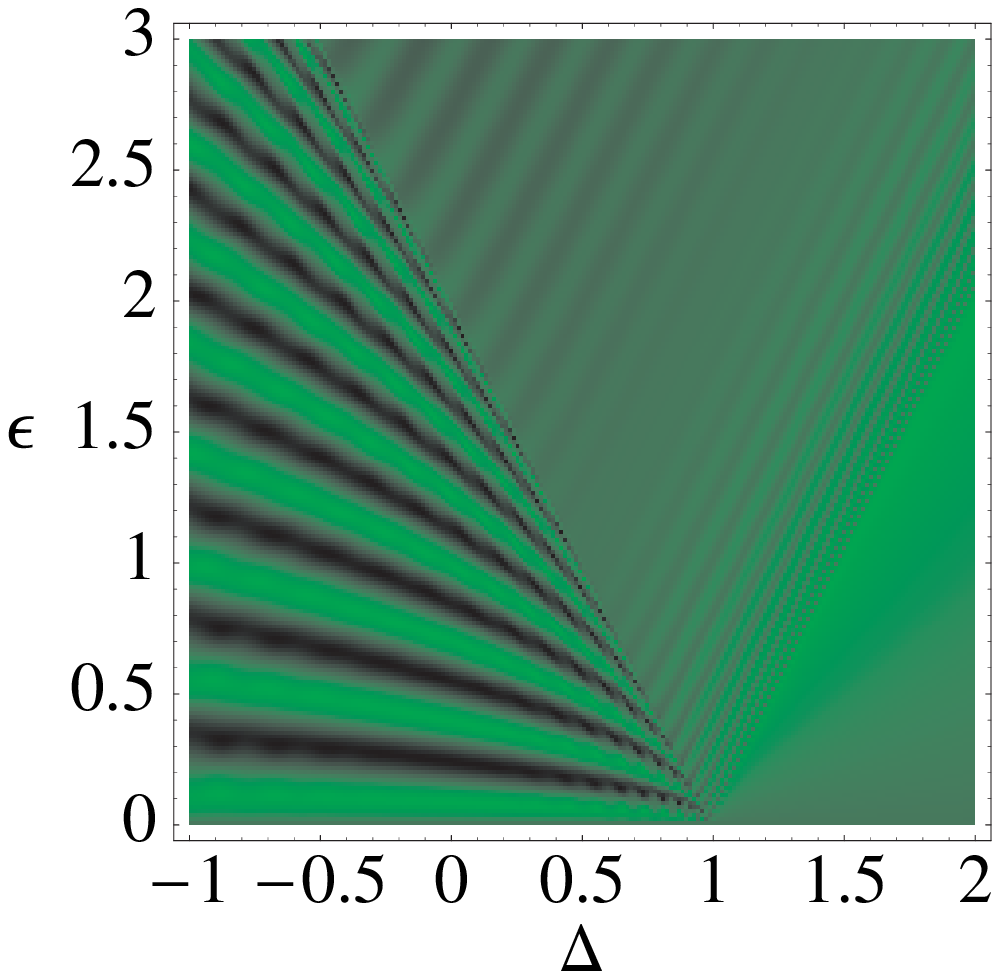}
(d)\\
\includegraphics[width=2in]{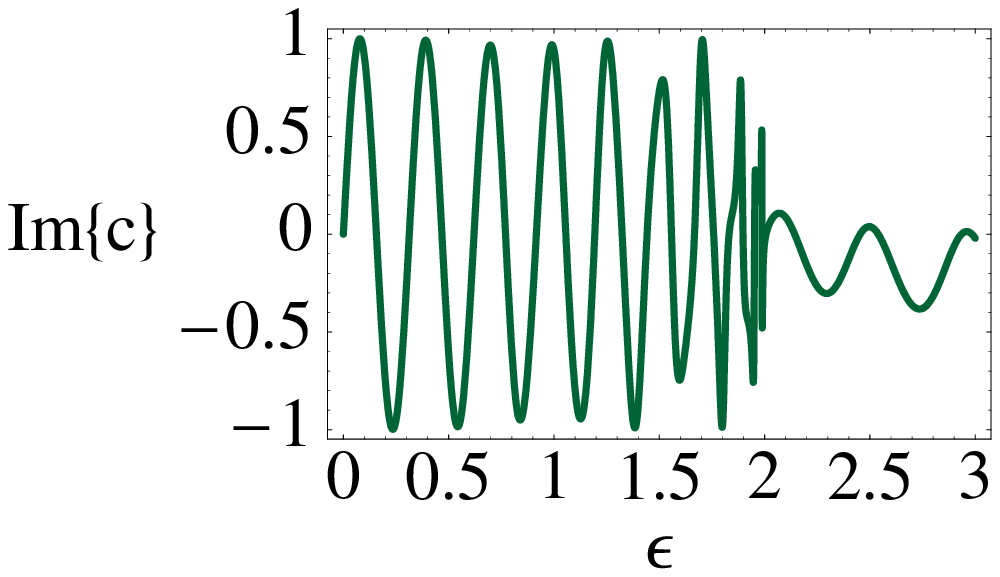}
\end{minipage}\hspace{.2in}
\begin{minipage}[b]{1.85in}
\caption{\label{fig-ReImc}\newline Real part~[(a),~(b)] and imaginary part~[(c),~(d)] of $c=1-t_+^*t_--r_+^*r_-$ as a function of the dimensionless exchange splitting~$\varepsilon$ and band shift~$\Delta$. It is computed for a multilayer structure characterized by~${k_\mathrm{in}w=40}$. The graphs~(b) and~(d) show a cut of the density plots~(a) and~(c) along the line~$\Delta=0$.}
\end{minipage}
\end{center}
\end{figure}

\section{Dynamics of a single ferromagnet}

We now investigate the magnetization dynamics in a structure with a single ferromagnetic layer such as in \fref{fig-1FMgen}, with a spin-polarized current incoming from the left~($\chi_\mathrm{aL}=\chi_\mathrm{in}$) and none from the right~($\chi_\mathrm{aR}=0$). The charge current density is given by~$j_\mathrm{in}=ev_\mathrm{in}\chi_\mathrm{in}^\dagger\chi_\mathrm{in}$. If the injected current is perfectly spin-polarized,~$\vec{n}_\mathrm{in}=\chi_\mathrm{in}^\dagger\hat{\vec{\sigma}}\chi_\mathrm{in}/(\chi_\mathrm{in}^\dagger\chi_\mathrm{in})$ is a unit vector pointing in the direction of the spin current density~\cite{note-spinpol}. Rewriting the equation of motion~$\rmd\vec{S}/\rmd t=\vec{F}_\mathrm{net}$ of the magnetization for the unit vector~$\vec{n}=\vec{S}/S$ and substituting the result~\eref{Eq-Fnet}, one obtains
\begin{equation}\label{Eq-EoMn}
\frac{\rmd\vec{n}}{\rmd t}=\Omega\left[-\Re\lbrace c\rbrace\vec{n}\times\left(\vec{n}\times\vec{n}_\mathrm{in}\right)+\Im\lbrace c\rbrace\vec{n}\times\vec{n}_\mathrm{in}\right].
\end{equation}
The time scale is set by the frequency~$\Omega=(\hbar/2S)(j_\mathrm{in}/e)$. The properties of the ferromagnetic layer enter through the real and imaginary part of the coefficient~$c=1-t_+^*t_--r_+^*r_-$, which take values in the intervals~$0\le\Re\lbrace c\rbrace\le2$ and~$-1\le\Im\lbrace c\rbrace\le1$ [see \fref{fig-ReImc}]. This coefficient~$c=g^{\uparrow\downarrow}-t^{\uparrow\downarrow}$ is the difference of the reflection and transmission mixing conductances~\cite{TseRMP05}. It is interesting to notice that the second term of the torque on the right-hand side of~\eref{Eq-EoMn}, pointing in the direction of~$\vec{n}\times\vec{n}_\mathrm{in}$, does not appear in the quasi-classical result obtained in~\cite{SloJMMM96} and in the expression derived in~\cite{WaiPRB00} upon ensemble averaging within a Landauer-B\"uttiker scattering formalism. However, such a term is also obtained within magnetoelectric circuit theory~\cite{KovPRB06} or by solving diffusive transport equations~\cite{GmiPRL06}.

\begin{figure}
\begin{center}
\begin{minipage}[b]{2in}
\includegraphics[width=2in]{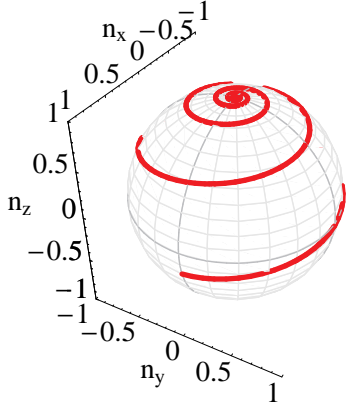}
\end{minipage}
\hspace{.2in}
\begin{minipage}[b]{4in}
\caption{\label{fig-soldyn1FM}The dynamics of the magnetization direction~$\vec{n}(t)$ starting from an initial condition~$\theta(0)=7\pi/12$ and~$\varphi(0)=0$, calculated with the parameter~$c=0.04-0.26\rmi$.}
\end{minipage}
\end{center}
\end{figure}

The equation of motion~\eref{Eq-EoMn} clearly conserves the unit length of~$\vec{n}$. It is convenient to use a coordinate system in which the $z$-axis points in the direction~$\vec{n}_\mathrm{in}$ of the spin polarization of the incoming current and to parametrize~$\vec{n}$ in spherical coordinates,~$\vec{n}=(\sin\theta\cos\varphi,\sin\theta\sin\varphi,\cos\theta)$. This yields the equations of motion
\numparts
\begin{eqnarray}
\rmd\theta/\rmd t&=& -\Omega\Re\lbrace c\rbrace\sin\theta(t)\;,\\
\rmd\varphi/\rmd t&=& -\Omega\Im\lbrace c\rbrace\;,  
\end{eqnarray}
\endnumparts
which have the solution
\numparts
\begin{eqnarray}
\tan\left[\theta(t)/2\right]&=&\exp\left(-\Re\lbrace c\rbrace\Omega t\right)\tan\left[\theta(0)/2\right],\\
\varphi(t)&=&-\Im\lbrace c\rbrace\Omega t+\varphi(0)\;, 
\end{eqnarray}
\endnumparts
for a given initial condition~$\theta(0),\ \varphi(0)$. Thus, in this simple model, the magnetization will align itself with the direction~$\vec{n}_\mathrm{in}$ of the spin polarization of the incoming current in a characteristic time~$(\Re\lbrace c\rbrace\Omega)^{-1}$, unless it initially points in the unstable anti-aligned direction~$-\vec{n}_\mathrm{in}$. This motion will be accompanied by a precession at frequency~$\Im\lbrace c\rbrace\Omega$ around the same direction~$\vec{n}_\mathrm{in}$. An example is shown in \fref{fig-soldyn1FM}.
%
%

Here we have investigated the dynamics induced by the spin-transfer torque only. A more complete analysis would take into account as well the effects of Gilbert damping and of a finite anisotropy field, as done \eg\ in~\cite{SloJMMM96,SunPRB00,GmiPRL06}. In this case, one typically finds either a steady precession of the magnetization or a magnetization reversal above a given current threshold, depending on the direction of the anisotropy field with respect to~$\vec{n}_\mathrm{in}$.

Realistic estimates for the parameters involved can be extracted from state-of-the-art experiments. In~\cite{KriSci05}, a magnetization density~$\mu_0 M_S=0.81\ \mathrm{T}$ has been reported for a 4-nm thick permalloy film, yielding~$S=\hbar M_S V/(g\mu_\mathrm{B})\approx 10^{6}\hbar$ for a layer of cross-section~$130\times60\ \mathrm{nm}^2$. This gives~$\Omega\approx3\ \mathrm{ns}^{-1}$ for an injected current~$j_\mathrm{in}=1\ \mathrm{mA}$. The band structure of a 3.5-nm thick permalloy film grown on Ni has been investigated by angle-resolved photoemission in~\cite{PetAPL98}. From this data we estimate~$\varepsilon\approx0.3$, assuming~$\Delta=0$ and taking~$k_\mathrm{in}\approx10\ \mathrm{nm}^{-1}$ in Cu. This yields~$c=0.04-0.26\rmi$ for~$w=4\ \mathrm{nm}$.

\section{Correlated dynamics of two ferromagnets}

\begin{figure}[t]
\begin{center}
\begin{minipage}[b]{3in}
\includegraphics[width=3in]{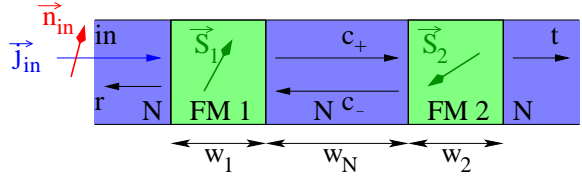}
\end{minipage}\hspace{0.2in}
\begin{minipage}[b]{3.1in}
\caption{\label{fig-2FM}A pillar with two ferromagnetic layers. A current with spin polarization along~$\vec{n}_\mathrm{in}$ is injected from the left. The component which is transmitted to the central metallic region can be reflected back and forth and induce correlations of the magnetizations~$\vec{S}_1$ and~$\vec{S}_2$ of the two ferromagnets.}
\end{minipage}
\end{center}
\end{figure}

The formalism presented in \sref{sec-STT1FM} can be applied to investigate the magnetization dynamics induced by the flow of a spin-polarized current in a system with two ferromagnetic layers such as the one depicted in \fref{fig-2FM}.
%
%
One component of the injected current is reflected by the first ferromagnet and another one is transmitted in the central region. There, it can be reflected back and forth before being transmitted to either side. Thus, the net spin fluxes through the two ferromagnets are correlated. As a consequence, the dynamics of the two magnetizations will also be correlated.

This effect can be investigated by solving the coupled equations of motion
\numparts
\begin{eqnarray}
\rmd\vec{n}_1/\rmd t&=&F_\mathrm{net}^{(1)}\;,\label{Eq-EoMn1}\\
\rmd\vec{n}_2/\rmd t&=&F_\mathrm{net}^{(2)}\;.\label{Eq-EoMn2}
\end{eqnarray}
\endnumparts
The net spin flux through the first ferromagnet~$\vec{F}_\mathrm{net}^{(1)}$ is given by substituting the magnetization~$\vec{n}_1$, the transmission and reflection amplitudes~$t_\pm^{(1)}$ and~$r_\pm^{(1)}$, and the incoming current amplitudes~$\chi_\mathrm{aL}=\chi_\mathrm{in}$ and~$\chi_\mathrm{aR}=\chi_\mathrm{c-}$ in expression~\eref{Eq-Fnet}. Similarly, the flux~$\vec{F}_\mathrm{net}^{(2)}$  is given in terms of~$\vec{n}_2$,~$t_\pm^{(2)}$, and~$r_\pm^{(2)}$, with the substitutions~$\chi_\mathrm{aL}=\chi_\mathrm{c+}$ and~$\chi_\mathrm{aR}=0$~(see \fref{fig-2FM}). Here again, by solving the scattering problem one can express the spinors in the central region in terms of the spinor of the incoming current,
\numparts
\begin{eqnarray}
\chi_\mathrm{c+}&=&\rme^{-\rmi k_\mathrm{in} w_1}\left[\hat{1}-\rme^{2\rmi k_\mathrm{in} w_\mathrm{N}}\hat{r}_1\hat{r}_2\right]^{-1}\hat{t}_1\chi_\mathrm{in}\;,\\
\chi_\mathrm{c-}&=&\rme^{\rmi k_\mathrm{in}[w_\mathrm{N}-(w_1+w_2)/2]}\hat{r}_2\left[\hat{1}-\rme^{2\rmi k_\mathrm{in} w_\mathrm{N}}\hat{r}_1\hat{r}_2\right]^{-1}\hat{t}_1\chi_\mathrm{in}\;.
\end{eqnarray}
\endnumparts
The transmission and reflection amplitudes of each ferromagnetic layer depend on the magnetization direction. This is the origin of the coupling between the equations of motion~\eref{Eq-EoMn1} and~\eref{Eq-EoMn2}. The problem becomes easier when the first ferromagnetic layer is much thicker than the second one, implying~$S_1\gg S_2$. Then,~$\vec{n}_1$ can be considered as fixed while solving the equation of motion for~$\vec{n}_2$. The solution of these problems is the object of work in progress.

In conclusion, we have evaluated the spin-transfer torque within a scattering approach for a single channel in the ballistic regime, and discussed the extension of this formalism to investigate correlated magnetization dynamics in multilayer magnetic nanopillars. 
\section*{Acknowledgments}
This work has been financially supported by the Swiss SNF and the NCCR Nanoscience.


\begin{thebibliography}{99}
\bibitem{note-spinpol} If the incoming current is not perfectly spin-polarized, the pure-state wave-function~$\chi_\mathrm{aL}\rme^{\rmi k_\mathrm{in} x}$ should be replaced by the appropriate mixture. In all fomulae,~$\vec{n}_\mathrm{in}$ should be replaced by a vector of length equal to the amount of spin polarization~$0\le p_\mathrm{in}\le1$ and pointing in the direction of the spin current density.
\end{thebibliography}
\end{document}